\documentclass[aps,twocolumn,showkeys,preprintnumbers,amsmath,amssymb,superscriptaddress,10pt]{revtex4-1}
\usepackage{graphicx}
\usepackage{dcolumn}
\usepackage{bm}
\usepackage{amsthm}
\usepackage{amsmath}
\usepackage{amssymb}
\usepackage{epstopdf}
\usepackage{color}  
\usepackage{float}

\begin{document}

\title{Single-stage direct Langevin dynamic simulations of transitions over arbitrary high energy barriers: Concept of the energy-dependent temperature}

\author{Dmitry Berkov}
\author{Elena K. Semenova}
\author{Natalia L. Gorn}

\affiliation{General Numerics Research Lab, Leutragraben 1, D-07743 Jena, Germany}

\begin{abstract}
In this paper we present an algorithm which allows {\it single-stage} direct Langevin dynamics simulations of transitions over {\it arbitrary high} energy barriers employing the concept of the energy-dependent temperature (EDT). In our algorithm, simulation time required for the computation of the corresponding switching rate {\it does not increase} with energy barrier. This is achieved by using in simulations an effective temperature which depends on the system energy: around the energy minima this temperature is high and tends towards the room temperature when the energy approaches the saddle point value. Switching times computed via our EDT algorithm show an excellent agreement with results obtained with the established forward flux sampling (FFS) method. As the simulation time required by our method does not increase with the energy barrier, we achieve a very large speedup when compared even to the highly optimized FFS version. In addition, our method does not suffer from stability problems occurring in multi-stage algorithms (like FFS and 'energy bounce' methods) due to the multiplication of a large number of transition probabilities between the interfaces.
\end{abstract}


\maketitle


\section{Introduction}
\label{Intro}

Evaluation of escape rates $\Gamma$ (or, equivalently, switching times $\tau_{\rm sw}$) over high energy barriers is a highly important and in most cases a very difficult task arising in any scientific area where systems with more than one stable states are studied - in physics, chemistry, molecular biology, material science etc. \cite{Haenggi_RevModPhys_1990}. This problem is much more difficult than the computation of the height of the corresponding energy barriers separating these metastable energy minima, because system dynamics near the saddle point may be highly non-trivial. For solution of the latter problem, several meanwhile standard methods have been implemented in the recent decades. The most widely used algorithm for this purpose is undoubtedly the 'nudged elastic band' (NEB) method of Jonsson et al. \cite{Jonsson_1998}, which employs the idea that the energy gradient component perpendicular to the optimal path should be zero along the whole path. The main advantage of NEB is the suggestion to connect the neighbouring system states along the transition path with artificial 'springs' to prevent a too large distance between these states during the path-finding procedure. Some less known methods are the closely related 'string method' which also searches for the 'minimal energy path', but in a slightly different way 
\cite{E_2002, Berkov_HMMM_2007} and the minimization of the Onsager-Machlup functional \cite{Onsager_1953}, first implemented by us for an interacting system of single-domain particles in \cite{Berkov_1998}. 

Obviously, to compute the average lifetime of a system with several metastable states - which is the quantity of interest for applications one needs more than the value of the energy barrier $\Delta E$. Even in the simplest analytical approximation for $\Gamma$ given by the Arrhenius law $\Gamma = \nu_{\rm att} \exp(\Delta E/k_{\rm B} T)$, the 'attempt frequency' $\nu_{\rm att}$, usually interpreted as the oscillation frequency near the metastable state is present. Omitting the discussion about the highly non-trivial task of computing this frequency for systems with internal degrees of freedom (see, e.g., \cite{Haenggi_RevModPhys_1990, Braun_1994, Fiedler_2012} etc.), we recall that the Arrhenius law is fundamentally not a satisfactory approach 
\cite{Kramers_1940, Coffey_Kalmykov_2012}, because the Arrhenius expression does not contain the system damping, which presence in the escape required by the fluctuation-dissipation theorem, as switching can occur only due to the interaction with the thermal bath.

The best possible analytical solution for the escape rate in a system with arbitrary damping (known as the Kramers problem) was derived in the famous paper of Mel'nikov and Meshkov \cite{Melnikov_1986}; this solution includes the intermediate-to-high damping (IHD) regime studied by Brown \cite{Brown_1979} and the very low damping (VLD) considered by Klik and G\"unther \cite{Klik_1990}. The formalism developed in \cite{Melnikov_1986} was successfully applied to a escape rate out of a single well and transition rates between two energy minima for a single-domain magnetic particle in \cite{Coffey_2001} and \cite{Dejardin_2001}; for the corresponding detailed review see \cite{Coffey_Kalmykov_2012}.

However, as any analytical approach, the expression given in \cite{Coffey_2001} has serious shortcomings. The most important one is that the analytical treatment is impossible for  magnetic particles with the size larger than the characteristic micromagnetic length \cite{Hubert_book_1998}, because magnetization configuration of this particles is spatially non-homogeneous. But even for the simplest case of single-domain particles, analytical methods cannot account for the 'back-hopping' trajectories (i.e., trajectories which, after crossing the saddle point, return to the starting minimum {\it before} they reach the equilibrium in the target minimum. Hence general numerical methods for the evaluation of the actual escape rate, which would take into account all the features listed above are strongly desired. 

Among these methods, the Langevin dynamics (LD) is conceptually the simplest one, because it directly mimics the time evolution of the system under the influence of thermal fluctuations. Unfortunately, LD is suitable for small barriers only ($\Delta E / k_B T \le 10$), because switching times (and correspondingly - the computation time) grow exponentially with $\Delta E$.

Thus, methods for evaluating numerically the escape rate over high barriers are usually based on a kind of gradual 'climbing' towards the saddle point uphill the energy surface. 

The most successful general method of this class is the so called forward flux sampling (FFS) \cite{Allen_2005, Allen_2006b, Borrero_2008, Allen_2009}. In FFS, the phase space between the two energy minima of interest is first divided into a (large) number of interfaces. Then the probability $w(\lambda_i \to \lambda_{i+1})$ to reach the next interface starting from the previous one is computed. In order to ensure that this probability is computed reasonably fast and accurately by standard LD simulations,  subsequent interfaces are placed relatively close to each other. Finally, multiplying the product of all these transition (for all interface pairs between the two minima) by the flux from the starting minimum through the first interface, one obtains the transition rate.

The interfaces are usually defined in the system coordinate space, using the sequence of values of the 'reaction coordinate', which defines whether the transition has occurred or not. In micromagnetics, this method was applied for magnetization switching in columnar recording structures (order parameter being the average magnetization projection) \cite{Vogler_2013, Vogler_2015, Desplat_2020b} and in skyrmions (order parameter was the skyrmion size) \cite{Desplat_2020a}. 

Computational time for FFS is roughly proportional to the energy barrier height, because for larger barriers more interfaces are needed in order to maintain the transition probabilities between the neighbouring interfaces reasonably high. However, an additional (and often really substantial) time effort is required for the optimal positioning of interfaces. This optimal positioning should ensure that transition probabilities 
$w(\lambda_i \to \lambda_{i+1})$ are the same for all interface pairs, because in this case the most accurate estimation of the transition rate is achieved \cite{Borrero_2008, Allen_2009}. Corresponding optimal placement required an iterative procedure which naturally requires several evaluations of the whole set of these probabilities, i.e., several complete FFS runs. We could demonstrate \cite{Semenova_Berkov_2020} that this large additional effort can be avoided if the interfaces are placed directly in the energy space, so that all probabilities $w_{i \to i+1}$ (which are $\sim \exp(-(E_{i+1}-E_i) / k_B T)$ are approximately equal.

Another inherent problem of FFS and related multi-stage climbing methods (e.g. the 'energy bounce' (EnB) algorithm \cite{Wang_Visscher_2006}) is the tight requirement to the accuracy of the numerically computed transition probabilities $w_i$ (evaluated by LD simulations). This accuracy should be really high because the final result includes the product of these probabilities so that any bias of $w_i$ will be elevated to the corresponding degree. Systematic errors are especially dangerous - it is easy to estimate that for a system with 50 interfaces, such an error of only 2\% in each $w_i$ would lead to the error of nearly 300\% in the final result. Even the stochastic mean-square error of only 5\% on each interface - a very low value for this kind of simulations - would lead to a relative error of $\approx 35\% $ in the computed switching rate.   

Thus, a new class of numerical methods which could perform the evaluation of the switching rate using only single-stage (in contrast to a gradual 'climbing' over a long series of interfaces as in FFS and EnB algorithms) Langevin dynamics simulations for energy barriers of arbitrary heights are highly desirable. In this study, we present such an algorithm based on the concept of the energy-dependent effective temperature. Our method allows stable and accurate single-stage simulations of transitions over any barrier with the simulation time which {\it does not increase} with the barrier height.

This paper is organized as follows: In Sec. \ref{Subsec_Main_idea} we describe the main idea of our algorithm: it is based on LD simulations of the system where the effective temperature depending on the system energy (EDT) is introduced: this temperature is high near the energy minima and tends to the room temperature in the vicinity of the saddle point(s). Then, in Section \ref{Sec_Ratio_EDT_Tconst} we derive the relation between the switching time obtained for the EDT system and the switching time of interest, i.e., for a constant temperature (CT). In the next Section \ref{Sec_MCh_formalism} the Markov chain used for the evaluation of the ratio of probability products for EDT and CT cases is constructed. Section \ref{Sec_EDT_validation} is devoted to the validation of our method via the EDT-version of the FFS algorithm. Finally, Sec. \ref{Sec_EDT_vs_FFS} contains the direct comparison of 'real' switching times obtained by EDT and standard ($T = Const$)  FFS algorithms. Here we show a very good agreement between both methods
in the energy barriers $10 \le \Delta E/ k_{\rm B} T \le 60$, where switching times span about 20 orders of magnitude. Further, we demonstrate a large speedup of the EDT algorithm as compared even to the optimized (as explained in \cite{Semenova_Berkov_2020}) FFS method.

%
\section{Energy-dependent temperature: methodology}
\label{EDT_method}

\subsection{Main idea}
\label{Subsec_Main_idea}

Direct LD simulations of transitions over high energy barriers are not feasible due to the major drawback of this method: the system spends the overwhelming majority of time in the vicinity of its energy minima, and the probability of approaching the landscape region near the saddle point is exponentially small ($p \sim \exp(-\Delta E / k_B T)$). 

To overcome this obstacle, we suggest to introduce the effective energy-dependent temperature $T(E)$. This temperature depends on the system energy in the following way: it is equal to the room temperature $T_{\rm room}$ 
for energies slightly below the saddle point 
($T(E) \to T_{\rm room}$ for $\Delta E - E \sim k_B T$), and is much higher than $T_{\rm room}$ for energies considerably lower than the energy barrier $\Delta E$: 
$T(E) = T_{\rm lrg} \gg T_{\rm room}$ for $\Delta E - E \gg k_B T$. 

\begin{figure}[htb]
\includegraphics[width=70mm]{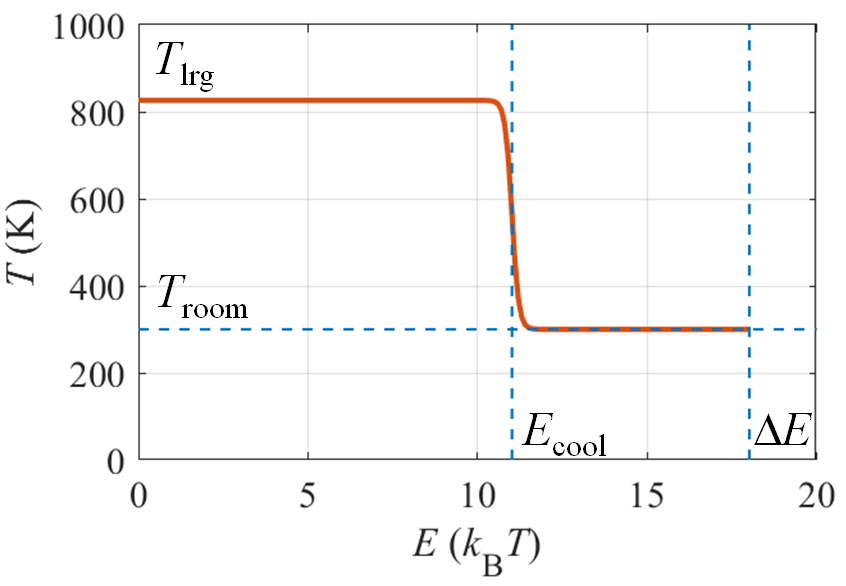}
\caption{Temperature as function of energy.}
\label{Fig_T_vs_E}
\end{figure}

For this purpose, we use the functional dependence
\begin{equation}
\label{Eq_T_vs_E}
T(E) = a_1 + a_2 \tanh 
\left( \frac {\Delta E - b_{\rm cool} \cdot k_B T}{\Delta_T} \right)
\end{equation}
(see Fig. \ref{Fig_T_vs_E}). The finite width $\Delta_T$ of this $T$-distribution should merely ensure a smooth transition between the 'hot' and 'cold' regions (abrupt temperature change would cause numerical instabilities of LD trajectories); we have checked that values $\Delta_T = (0.1 \div 1.0) k_B T$ lead to the same final results. Parameters $a_1$ and $a_2$ should be chosen to satisfy the two conditions 
\begin{equation}
\label{a1_a2_conditions}
\begin{split}
T(E: E - E_{\rm cool} \gg k_B T) \to T_{\rm room} \\
T(E: E_{\rm cool} - E \gg k_B T) \to T_{\rm lrg} 
\end{split}
\end{equation}
('cold' region near the barrier and 'hot' region far below the barrier), so that $a_1 = (T_{\rm room}+T_{\rm lrg})/2$ and $a_2 = (T_{\rm room}-T_{\rm lrg})/2$.

Temperature $T_{\rm lrg}$ itself and the 'cooling' energy 
$E_{\rm cool}  = \Delta E - b_{\rm cool} \cdot k_B T_{\rm room}$ 
should be set so that the probability 
$p(E_{\rm cool}) \sim \exp(-E_{\rm cool}/k_B T_{\rm lrg})$ 
to occupy states near $E_{\rm cool}$ is large enough to frequently provide 'launching points' for the system to overcome the energy barrier starting from this energy. Basing on desired values of $p(E_{\rm cool}) = 0.001 - 0.01$, we obtain 
$E_{\rm cool}/k_B T_{\rm lrg} = a_{\rm lrg} \sim 4 - 6$. We have used $a_{\rm lrg} = 4$; its further increase naturally led to fewer observed transitions and poorer statistics.

The last parameter to be determined - $b_{\rm cool}$ - controls the height of the effective energy barrier $\Delta E_{\rm eff} = \Delta E - E_{\rm cool} = b_{\rm cool} \cdot k_B T_{\rm room}$ which the system has to overcome starting from the energy $E_{\rm cool}$. The upper limit of $b_{\rm cool}$ is set by the ability to overcome the corresponding barrier employing standard LD simulations. On the other hand, too small vales of $b_{\rm cool}$ lead to very frequent crossings of the energy barrier, so that it is difficult to distinguish between 'true' and 'false' transitions between the basins (see \cite{Semenova_Berkov_2020} for the detailed discussion). These arguments lead to the parameter range $5 \le b_{\rm cool} \le 10$; in our simulations we have used mostly $b_{\rm cool} = 7$ and have checked that varying it in above mentioned limits does not change final results within the statistical accuracy.

An example of the dependence $T(E)$ with parameters given above is shown in Fig. \ref{Fig_T_vs_E} for a system with the energy barrier $\Delta E = 18 k_B T$.

It is clear that for the system with the EDT profile (1) we should observe numerous transitions over the barrier $\Delta E$ by employing direct LD simulations, no matter how large this barrier is: effective temperature for energies $E < \Delta E - b_{\rm cool} k_B T$ is high enough to ensure a significant occupation of these states, so that the energy barrier to be overcome is only $\Delta E^{\rm EDT} \simeq b_{\rm cool} k_B T$. The corresponding switching time for an EDT-system thus can be computed in a standard way using LD simulations, namely dividing the physical simulation time by the number of 'true' switchings: $\tau_{\rm sw}^{\rm EDT} = t_{\rm sim}/N_{\rm sw}$ \cite{Semenova_Berkov_2020}.

The key problem is how to establish the relation between this EDT-computed switching time $\tau_{\rm sw}^{\rm EDT}$ and the switching time for the same system at a constant temperature $\tau_{\rm sw}^{\rm CT}$ - a quantity of a real physical interest. 

\subsection{Relation between the EDT-computed time and the real switching time}
\label{Sec_Ratio_EDT_Tconst}

To establish the above mentioned relation, we start with the same expression for the transition rate $\Gamma$ (recall that $\tau_{\rm sw} = 1/\Gamma$) which is used in forward-flux sampling (FFS) algorithms: we introduce virtual interfaces 
$\{ \lambda_i, i=1,...,N \}$ between the basins {\bf A} and {\bf B} (whereby 
$\lambda_1 \equiv \lambda_{\bf A}$, $\lambda_N \equiv \lambda_{\bf B}$), so that 
\begin{equation}
\label{Eq_Trans_rate_general}
\Gamma_{{\bf A} \to {\bf B}} = 
\Phi_{\lambda_1 \to \lambda_2} \cdot \displaystyle \prod_{i=2}^{N-1}
w(\lambda_i \to \lambda_{i+1}) \equiv 
\Phi_{\bf A} \cdot \displaystyle \prod_{i=2}^{N-1} w_{i \to i+1}   
\end{equation}
Note  that we have slightly changed the numbering of interface compared to our previous paper \cite{Semenova_Berkov_2020} to make it consistent with the numbering of Markov chain states used in the next sections. 

Eq. (\ref{Eq_Trans_rate_general}) represents a very general statement that transition rate 
$\Gamma_{\rm {\bf A} \to {\bf B}}$ can be viewed as the product of the flux $\Phi_{\bf A}$ out of the basin {\bf A} through the interface $\lambda_2$ (i.e., the number of particles per unit time starting in {\bf A} and crossing the first interface outside {\bf A}), and the subsequent conditional probabilities $w(\lambda_i \to \lambda_{i+1})$ that a particle starting from the interface $i$ reaches the interface $i+1$.

Using Eq. (\ref{Eq_Trans_rate_general}), the ratio $\tau_{\rm sw}^{\rm CT} / \tau_{\rm sw}^{\rm EDT}$ can be written as
\begin{equation}
\label{Eq_Trans_rates_ratio}
\frac{\tau _{\rm sw}^{\rm CT}}{\tau _{\rm sw}^{\rm EDT}}=
\frac{\Gamma _{{\bf A} \to {\bf B}}^{\rm EDT}}
     {\Gamma _{{\bf A} \to {\bf B}}^{\rm CT}} =
\frac{\Phi_{\bf A}^{\rm EDT}(T=T_{\rm lrg})}{\Phi_{\bf A}^{\rm CT}(T=T_{\rm room})} \cdot 
\frac{\displaystyle \prod_{i=2}^{N-1} w_{i \to i+1}^{\rm EDT}}
     {\displaystyle \prod_{i=2}^{N-1} w_{i \to i+1}^{\rm CT}}
\end{equation}
where the initial fluxes should be computed at corresponding temperatures, as explicitly indicated in (\ref{Eq_Trans_rates_ratio}). Hence, the actual switching time is
\begin{equation}
\label{Eq_tau_sw_real}
\tau _{\rm sw}^{\rm CT} = \tau _{\rm sw}^{\rm EDT} \cdot 
\frac{\Phi_A^{\rm EDT}}{\Phi_A^{\rm CT}} \cdot 
\frac{\displaystyle \prod_{i=2}^{N-1} w_{i \to i+1}^{\rm EDT}}   
     {\displaystyle \prod_{i=2}^{N-1} w_{i \to i+1}^{\rm CT}}
\end{equation}
As explained above, $\tau_{\rm sw}^{\rm EDT}$ in this expression can be computed from direct LD simulations. The fluxes $\Phi_{\bf A}^{\rm EDT}$ and $\Phi_{\bf A}^{\rm CT}$ are also easily available from such simulations, because the first interface is usually chosen to be close ($\sim k_B T$) to the basin {\bf A}. Thus, our task reduces to the evaluation of the ratio of two probability products 
\begin{equation}
\label{Eq_ratio_probprod}
r = \frac 
{\displaystyle \prod_{i=2}^{N-1} w_{i \to i+1}^{\rm EDT}}
{\displaystyle \prod_{i=2}^{N-1} w_{i \to i+1}^{\rm CT}}
\end{equation}
We emphasize that the method for evaluation of this ratio should be either an analytical one or a numerical method with a very low computational effort, because otherwise the EDT algorithm will not have any advantage compared to standard FFS methods. In the next Section, we shall construct a Markov chain which enables the evaluation of (\ref{Eq_ratio_probprod}) using only $N$ diagonalizations of matrices with the sizes 
$\le N$.
%
\section{Markov chain formalism}
\label{Sec_MCh_formalism}

\subsection{Evaluation of the of the equilibrium probabilities $w_{i \to i+1}$ using Markov chains}
\label{Sec_Calc_prob_ratio}
\begin{figure}[htb]
\includegraphics[width=80mm]{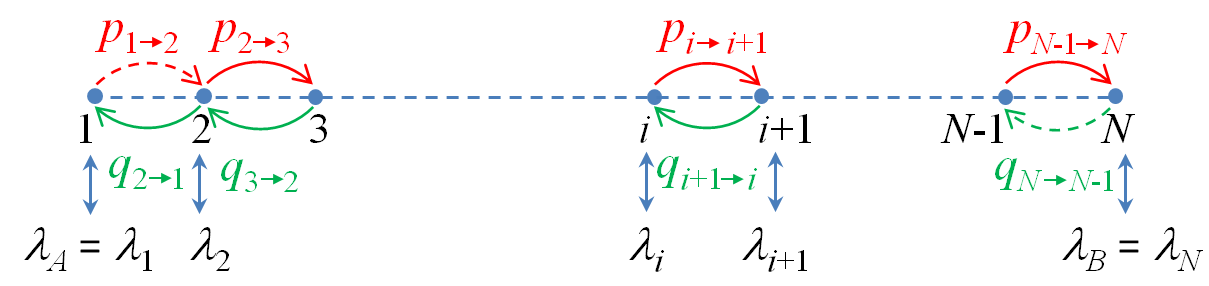}
\caption{Markov chain consisting of $N$ states $\{ 1, ..., N \}$ which correspondence to the interfaces $\{ \lambda_A, ..., \lambda_B \}$ is shown by black arrows.}
\label{Fig_MarkovChain_definition}
\end{figure}

In this subsection we demonstrate how to compute the required ratio (\ref{Eq_ratio_probprod}) of probability products using the Markov chain (MCh) formalism (see, e.g., \cite{Howard_DynProbSys_book}). For this purpose, we introduce the Markov chain with the set of states $\{ i = 1, ..., N \}$, which correspond to our set of interfaces 
$\{ \lambda_A = \lambda_1, ..., \lambda_i, ...,  \lambda_N = \lambda_B \}$. We denote the {\it one-step} transition probabilities between these chain states as $p_{i \to i+1}$ and $q_{i \to i-1}$. Corresponding Markov chain for the whole set of interfaces is shown in Fig. \ref{Fig_MarkovChain_definition}.

Probabilities $p_{i \to i+1}$ and $q_{i \to i-1}$ form the one-step transition matrix $\hat{\bf P}$ for the Markov chain (which governs the change of the state occupations in this chain after one step): $P_{i,i+1} = p_{i \to i+1}$ and $P_{i,i-1} = q_{i \to i-1}$.  

To compute the transition probabilities $w_{i \to i+1}$ appearing in the basic expression (\ref{Eq_Trans_rate_general}) (and correspondingly - in Eqs. (\ref{Eq_Trans_rates_ratio})-(\ref{Eq_ratio_probprod})), we first recall how these probabilities are defined: a system trajectory is started from the interface $\lambda_i$ and simulated (using the standard Langevin dynamics) until it either arrives at the next interface $\lambda_{i+1}$ or returns to the basin ${\bf A}$. Then the next trajectory is launched from $\lambda_i$ etc. Probability $w_{i \to i+1}$ is defined as the fraction of launched trajectories which arrive at $\lambda_{i+1}$.

According to this procedure, the random process for which we construct the Markov chain for the evaluation of $w_{i \to i+1}$, terminates when the system reaches either the state $1$ or the state $(i+1)$. Hence, we have the Markov chain of the length $(i+1)$ with absorbing borders, so that corresponding elements of the one-step transition matrix 
$\hat{\bf P}^{(i+1)}$ of this chain are $P^{(i+1)}_{11} = P^{(i+1)}_{i+1,i+1} = 1$, 
$P^{(i+1)}_{12} = p_{1 \to 2} = 0$ and $P^{(i+1)}_{i+1,i} = q_{i+1 \to i} = 0$. The whole matrix $\hat{\bf P}^{(i+1)}$ is then tridiagonal and has the form
\begin{equation}
\label{Eq_One-step_trans_mtr_partial}
    \hat{\bf P}^{(i+1)} =
    \begin{bmatrix}
    1 & 0 & &  & \cdots & & 0\\
    q_{21} & 0 & p_{23} & &  \\
    0 & q_{23} & 0 & p_{34} & &  \\
    \vdots &  && \ddots &  & & \vdots\\
    & & & & 0 & p_{i-1,i} & \\
    &  & & & q_{i,i-1} & 0 & p_{i,i+1}\\
    0 & & & &\cdots  & 0 & 1
    \end{bmatrix}
\end{equation}
This matrix belongs to the class of the so called stochastic matrices, for which the sum of elements of each row is one.

Next, we recall that $w_{i \to i+1}$ is computed from LD simulations which are carried out until the system reaches either the interface $\lambda_{i+1}$ or the basin {\bf A}, i.e., without restricting the simulation time. In the Markov chain formalism this corresponds to the probability that the system, being initially in the $i$-th state, will be found in the $(i+1)$-th state after an arbitrary large number of steps (equilibrium configuration). Thus, in order to compute $w_{i \to i+1}$ from the one-step matrix $\hat{\bf P}^{(i+1)})$, we have to find the matrix $\hat{\bf E}^{(i+1)} = \lim_{k \to \infty} (\hat{\bf P}^{(i+1)})^k$. The probability of interest is then given by the corresponding matrix element of $\hat{\bf E}^{(i+1)}$, namely $w_{i \to i+1} = E_{i,i+1}^{(i+1)}$.

Importantly, the matrix $\lim_{k \to \infty} (\hat{\bf P}^{(i+1)})^k$ can be computed very fast: after the diagonalization of the matrix 
$\hat{\bf P}^{(i+1)} = \hat{\bf Q} \hat{\bf D} \hat{\bf Q^{-1}}$ 
this limit becomes 
$\lim_{k \to \infty} (\hat{\bf P}^{(i+1)})^k = 
\lim_{k \to \infty} \hat{\bf Q} \hat{\bf D}^k \hat{\bf Q^{-1}}$, so that we have to evaluate only the limits $\lim_{k \to \infty} d_j^k$ for eigenvalues of the matrix $\hat{\bf P}^{(i+1)}$. According to the properties of stochastic matrices, all their eigenvalues obey the inequality $d_i \le 1$, so that corresponding limits are either 0 or 1. 

%
\subsection{Assignment of one-step probabilities $\{p\}$ and $\{q\}$}
\label{Subsec_One_step_prob}
To assign the one-step probabilities $p_{i\to i+1}$ and $q_{i+1 \to i}$ for the Markov chain, we have first to establish the correspondence between the energy landscape and these states in our case. In our previous paper \cite{Semenova_Berkov_2020} we have proposed to place the interfaces for FFS simulation of the transition ${\bf A} \to {\bf B}$, not according to the values of magnetic moment projections (as it is done usually \cite{Vogler_2013, Vogler_2015}), but equidistantly in the energy space of the studied system. This positioning has greatly simplified the FFS algorithm, because the probabilities $w_{i \to i+1}$ depend mainly on the energy differences between the interfaces ($w_{i \to i+1} \sim \exp(-(E_{i+1} - E_i)/kT)$). Hence, for energy-equidistant interfaces these probabilities should be approximately the same for all 'uphill' interface pairs $i \to i+1$, what should minimize the statistical error of FFS \cite{Borrero_2008, Allen_2009}.

\begin{figure}[htb]
\includegraphics[width=80mm]{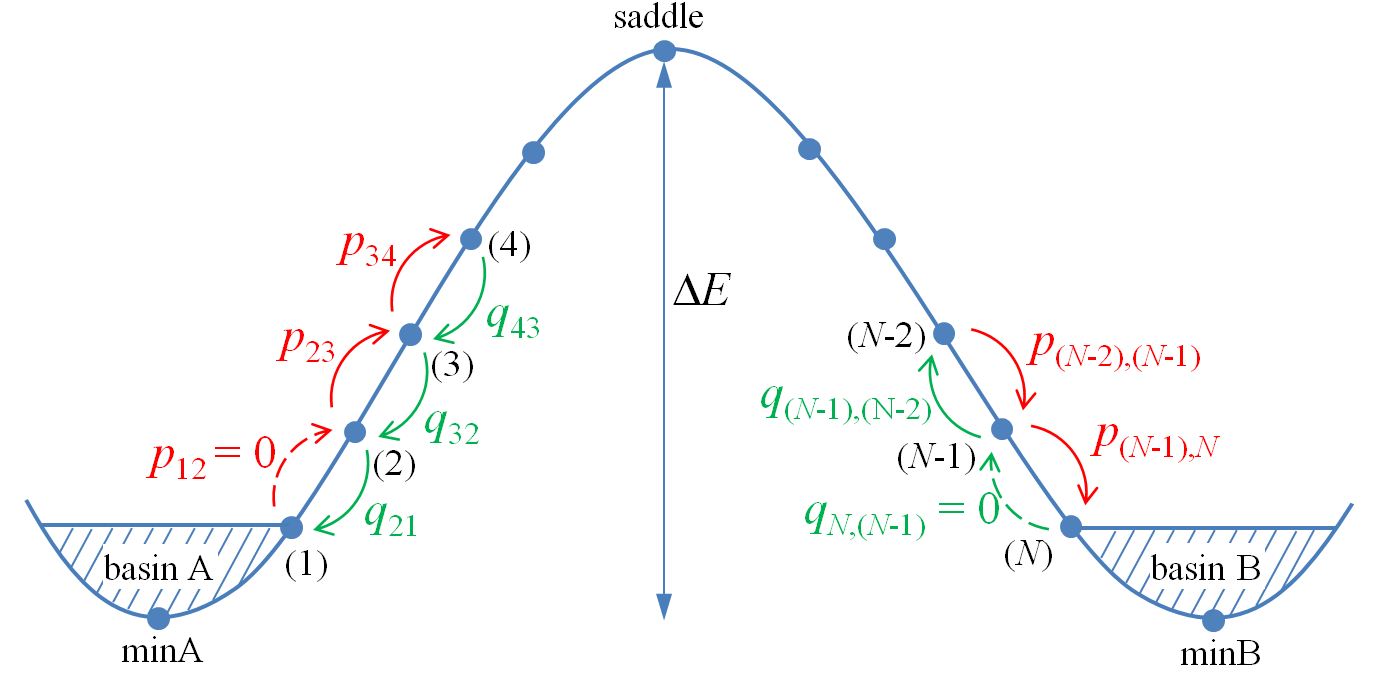}
\caption{Markov chain and energy landscape.}
\label{Fig_MarkovChain_EnLand}
\end{figure}

In our EDT algorithm presented here we use the same principle to position the interfaces and correspondingly - Markov chain states, as shown in Fig. \ref{Fig_MarkovChain_EnLand}.

This interface placement allows us to assign the MCh probabilities $p_{i\to i+1}$ and $q_{i+1 \to i}$ using the equilibrium thermodynamics and the principle of the detailed balance (see, e.g., \cite{vanKampen_book}). According to this principle, one-step MCh probabilities $p_{i \to j}$ and 
$q_{j \to i}$ are related to the equilibrium probabilities to find the system in the corresponding states $\pi_i$ and $\pi_j$ as $\pi_i p_{i \to j} = \pi_j q_{j \to i}$. Further, in a thermodynamic equilibrium these latter probabilities are given by 
$\pi_i \simeq n_i \exp(-E_i/k_B T)$, where $n_i$ is the density of states at the energy $E_i$. Hence, one-step MC probabilities $p_{i\to i+1}$ and $q_{i+1 \to i}$ should obey the relation

\begin{equation}
\label{Eq_p_over_q}
\frac{p_{i \to i+1}}{q_{i+1 \to i}} = \frac{\pi_{i+1}}{\pi_i} = 
\frac{n_{i+1} e^{-E_{i+1}/k_B T}} {n_i e^{-{E_i}/k_B T}} = 
\frac{n_{i+1}}{n_i} \exp\left(-\frac{\delta E_{i,i+1}}{k_B T}\right)    
\end{equation}
where $\delta E_{i,i+1} = E_{i+1} - E_i$.

To satisfy this relation, we set
\begin{equation}
\label{Eq_p_MCprob}
p_{i \to i+1} = \left( \frac{n_{i+1}}{n_i} \right)^{1/2}
               \exp \left( -\frac{1}{2} \frac{\delta E_{i,i+1}}{k_B T} \right)
\end{equation}
\begin{equation}
\label{Eq_q_MCprob}
q_{i+1 \to i}=\left( \frac{n_i}{n_{i+1}} \right)^{1/2}
    \exp \left( +\frac{1}{2} \frac{\delta E_{i,i+1}}{k_B T} \right)
\end{equation}
To evaluate the ratio $n_{i+1}/n_i$ for two subsequent states we note that for a small energy increments $\delta E_{i,i+1} \equiv \delta E$ we can expand $n_{i+1}=n(E_{i+1})$ into the Taylor series near $E = E_i$, obtaining
\begin{equation}
\label{Eq_DoS_expansion}
n_{i+1} = n_i + \left. \frac{\partial n}{\partial E} \right|_{E = E_i} \delta E = 
          n_i \left( 1 + \frac{\delta E}{n_i} 
          \left. \frac{\partial n}{\partial E} \right|_{E=E_i} \right)
\end{equation}
so that the required ratio is
\begin{equation}
\label{DoS_ratio}
{\left( \frac{n_{i+1}}{n_i} \right)^{\pm 1/2}} = 
1 \pm \frac{\delta E}{2 n_i} \left. \frac{\partial n}{\partial E} \right|_{E = E_i}
\end{equation}
Thus, for energies where the function $n(E)$ is non-singular (what is normally the case if $E$ does not correspond to an extremum of a saddle point) we can set 
$n_{i+1}/n_i \approx 1$ for small $\delta E \to 0$. Finally, we have to normalize $p$'s and $q$'s so that $p_{i \to i+1} + q_{i \to i-1} = 1$ to satisfy the normalization condition for transition probabilities out of the state $i$. 

\section{Validation of the EDT algorithm}
\label{Sec_EDT_validation}

\begin{figure}[htb]
\includegraphics[width=80mm]{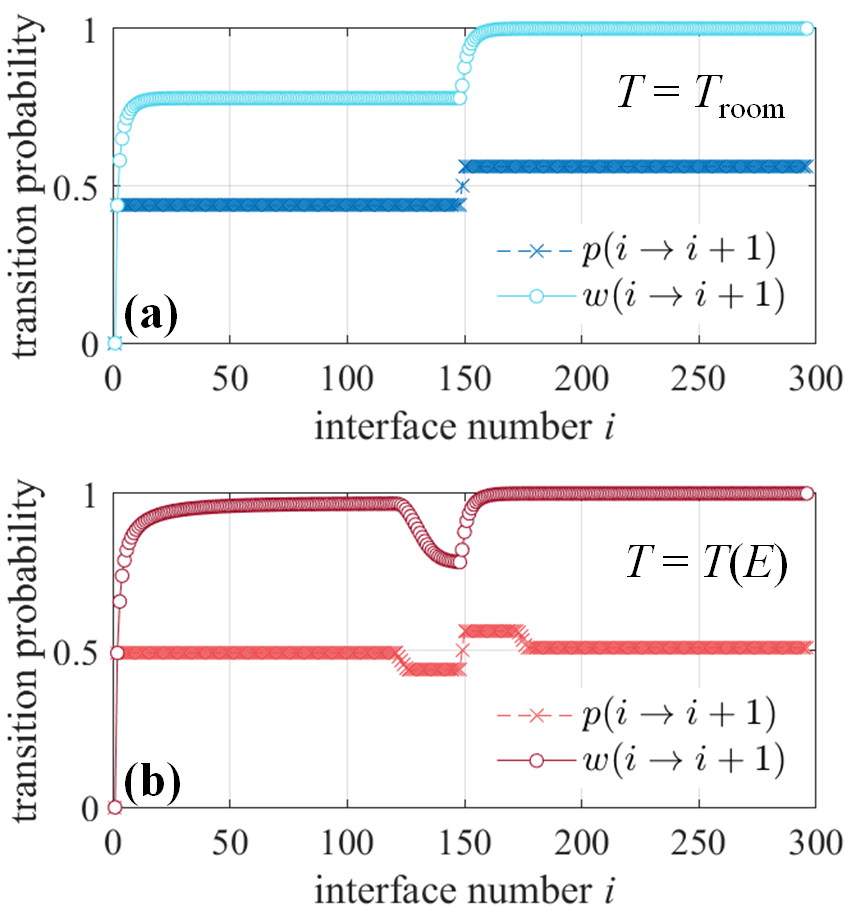}
\caption{One-step probabilities $p_{i \to i+1}$ (lines with crosses) and total probabilities $w_{i \to i+1}$ (lines with circles) as functions of the interface number for $T = T_{\rm room}$ (a) and $T = T(E)$ (b).}
\label{Fig_trans_prob_p_and_w}
\end{figure}

Dependencies of probabilities $p_i$ on the interface number $i$ for the whole Markov chain are shown in Fig. \ref{Fig_trans_prob_p_and_w} ($q_i = 1 - p_i$ and thus are not shown) for $T = T_{\rm room} = Const$ and $T = T(E)$ (1) as lines marked with crosses; in this example here the barrier is $\Delta E = 38 k_B T$ and the interface distance 
$\delta E = 0.25 k_B T$. 

According to the definition (\ref{Eq_p_MCprob}), for the case $T = Const$ the one-step probabilities $p_i$ should exhibit a jump for the interface corresponding to the saddle point (i.e., the middle interface, see Fig. \ref{Fig_trans_prob_p_and_w}(a)), because at this point the energy difference $E_{i+1} - E_i$ changes its sign. The same Eq. (\ref{Eq_p_MCprob}) implies that for the energy-dependent temperature (see Fig. \ref{Fig_T_vs_E}), the values of $p_i$'s should rapidly change also around the interface corresponding to the energy $E_{\rm cool} = \Delta E - b_{\rm cool} k_B T$ (line with crosses in Fig. \ref{Fig_trans_prob_p_and_w}(b)), where the temperature drops from $T_{\rm lrg}$ to $T_{\rm room}$.



Total transition probabilities $w_{i \to i+1}$ obtained from these one-step quantities as explained above (i.e., as $w_{i \to i+1} = E_{i,i+1}^{(i+1)}$, where 
$\hat{\bf E}^{(i+1)} = \lim_{k \to \infty} (\hat{\bf P}^{(i+1)})^k$ with the matrix $\hat{\bf P}$ given by (\ref{Eq_One-step_trans_mtr_partial})), are shown in the same Fig. \ref{Fig_trans_prob_p_and_w} as lines marked by circles. It can be seen that after the jump of $p_i$ the total probability $w_{i \to i+1}$ changes {\it smoothly}, tending to its new limit for the new constant temperature: 
$w_{i \to i+1} \to \exp(-\delta E/k_B T)$ for $\delta E > 0$ \cite{Semenova_Berkov_2020}. This behaviour is in accordance with the physical sense of the quantity $w_{i \to i+1}$ defined as the result of an unlimited number of steps for the Markov chain with the matrix  (\ref{Eq_One-step_trans_mtr_partial}). For example, it is clear that the total probability $w_{i \to i+1}$ to reach the next interface - i.e., not to return back to the basin 
{\bf A} - should gradually increase when the distance to this basin increases.

\begin{figure}[htb]
\includegraphics[width=80mm]{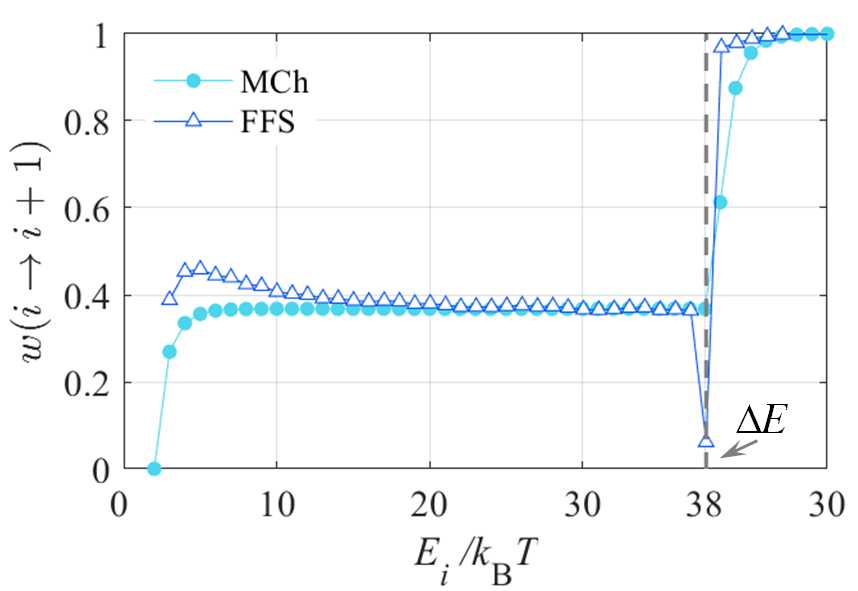}
\caption{Total transition probabilities $w_{i \to i+1}$ for FFS (open triangles) and MCh (closed circles) methods for the constant temperature $T = 300\,K$ (system with $\Delta E/k_B T = 38$ and $\delta E/k_B T = 1$).}
\label{Fig_prob_FFS_vs_MCh_Tconst}
\end{figure}

\begin{figure}[htb]
\includegraphics[width=80mm]{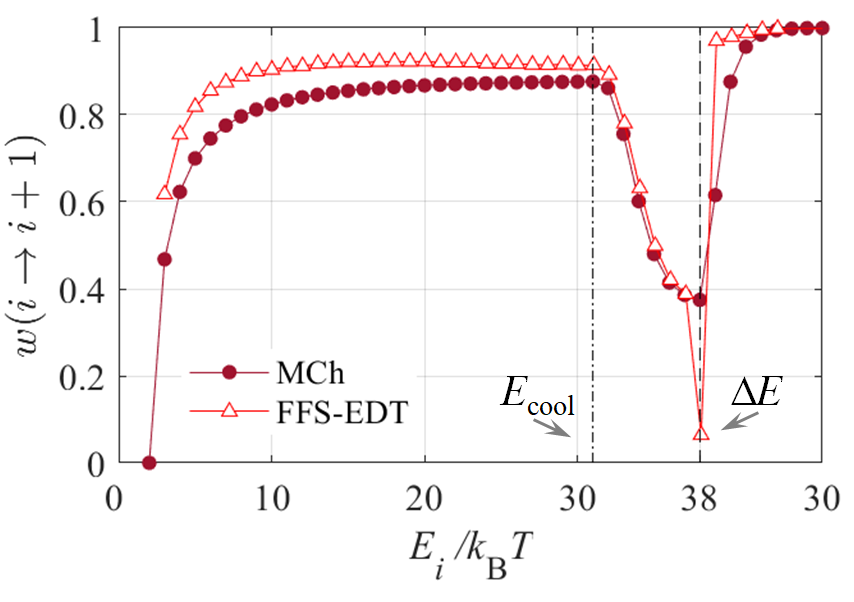}
\caption{The same as in Fig. \ref{Fig_prob_FFS_vs_MCh_Tconst} for the energy-dependent temperature T(E) given by (\ref{Eq_T_vs_E}).}
\label{Fig_prob_FFS_vs_MCh_EDT}
\end{figure}

The behaviour of the Markov chain probabilities $w_{i \to i+1}^{\rm MCh}$ can be better understood by comparing them to the same quantities calculated with the FFS which uses the same energy-equidistant interfaces (see \cite{Semenova_Berkov_2020} for the details of the latter method). Results of this comparison for the system with $\Delta E = 38 k_B T$ and interface distance $\delta E = 1 k_B T$ are shown for the case $T = Const$ in Fig. \ref{Fig_prob_FFS_vs_MCh_Tconst} and for the energy-dependent temperature $T(E)$ (\ref{Eq_T_vs_E}) - in Fig. \ref{Fig_prob_FFS_vs_MCh_EDT}.


\begin{figure}[htb]
\includegraphics[width=80mm]{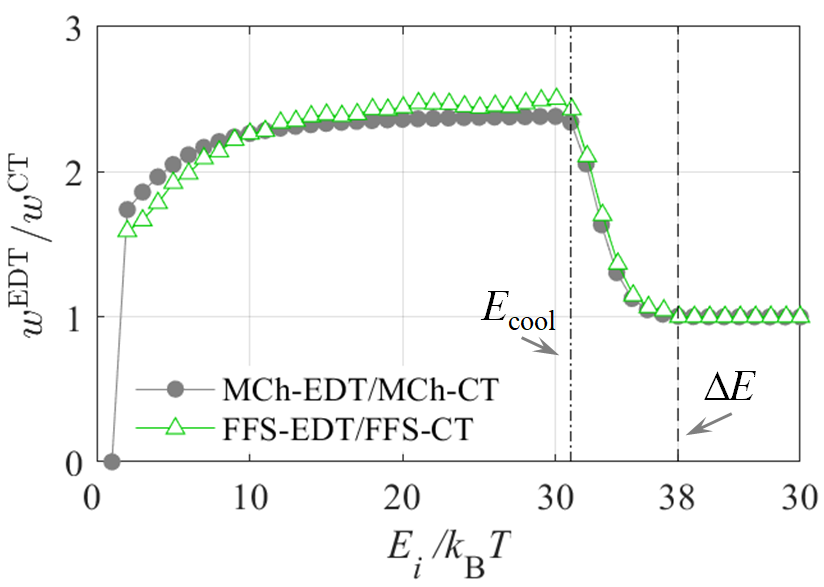}
\caption{Ratio of probabilities $w_{i \to i+1}^{\rm EDT}/w_{i \to i+1}^{T = Const}$ for FFS (open triangles) and MCh (full circles) methods as function of the interface energy. A very good agreement of these ratios for both methods is clearly demonstrated.}
\label{Fig_Prob_w_ratios_EDT_to_Tc}
\end{figure}

First we note that both for $T = Const$ and EDT the difference between $w_{i \to i+1}$ obtained by MCh and FFS for interface energies below the saddle point is due to the rather large value of the interface distance ($\delta E = 1 k_B T$) used here, so that neglecting the change in the density of states $n(E)$ in the MCh method according to the expansion (\ref{DoS_ratio}) has a noticeable effect. However, this difference decreases with $\delta E \to 0$, as explained above. 

\begin{figure}[htb]
\includegraphics[width=80mm]{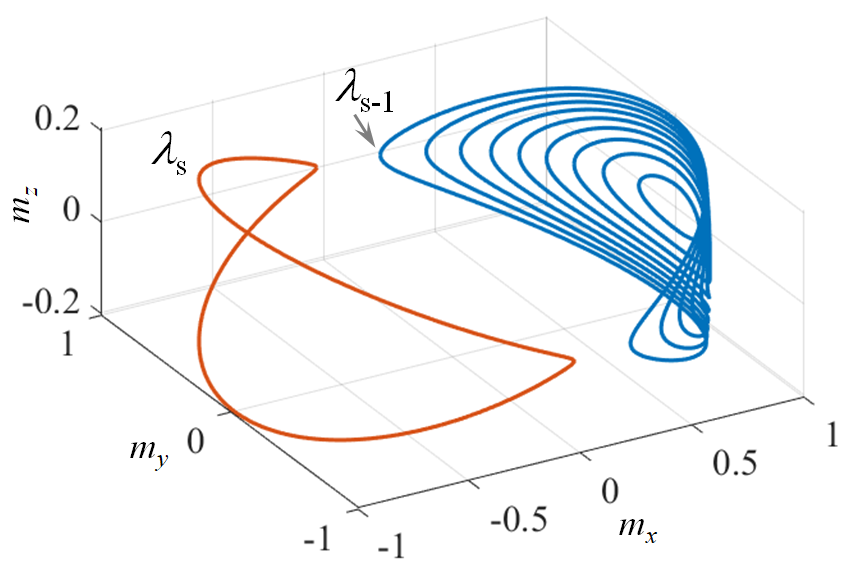}
\caption{To the explanation of the dip on the dependencies $w_{i \to i+1}(E_i)$ in Figs. \ref{Fig_prob_FFS_vs_MCh_Tconst} and \ref{Fig_prob_FFS_vs_MCh_EDT}: the interface $\lambda_{\rm s-1}$ immediately before the saddle point and the next (saddle) interface $\lambda_{\rm s}$ are separated by a large distance in the coordinate space.}
\label{Fig_Intf_3D}
\end{figure}

The most important feature of $w_{i \to i+1}$ seen in Fig. \ref{Fig_prob_FFS_vs_MCh_Tconst} and \ref{Fig_prob_FFS_vs_MCh_EDT} is the large discrepancy between MCh and FFS probabilities  at and slightly above the saddle point energy $\Delta E$. This discrepancy reflects the qualitative difference between the FFS and MCh methods. Namely, in FFS we evaluate $w_{i \to i+1}$ by LD simulations taking into account complicated physical processes near the saddle point (back-hopping in the first place) and the peculiarities of FFS interfaces for the specific system under study. In particular, for our macrospin the probability to reach the saddle-point interface is especially small (large dips at 
$E_i = \Delta E$ on $w_i$-dependencies for FFS), because the conditions to reach this interface also include the requirement that $m_x$-projection changes its sign (see \cite{Semenova_Berkov_2020} for details). For this reason the distance between the saddle interface and the previous one in the coordinate space is much larger than for preceding interface pairs (see Fig. \ref{Fig_Intf_3D}), leading to the correspondingly small probability $w(\lambda_{\rm s-1} \to \lambda_{\rm s})$. For the same reason, 
$w_{i \to i+1}$ strongly increases immediately after this interface, because the probability to return to previous interfaces is very low. In contrast, in MCh we merely compute the limit $\hat{\bf E}^{(i+1)} = \lim_{k \to \infty} (\hat{\bf P}^{(i+1)})^k$ where one-step probabilities $p_i$ and $q_i$ in the matrix $\hat{\bf P}$ have only a relatively small jump near the saddle point, so that $w_{i \to i+1}$ changes in the vicinity of the saddle point much slower than for FFS.

However - and this is the key point of our method - the {\it ratio} of probabilities $w$ for constant and energy-dependent temperatures
$w_{i \to i+1}^{\rm EDT}/w_{i \to i+1}^{\rm CT}$ should be the same (in the limit $\delta E \to 0$) in both FFS and MCh methods for all interface energies, including the region near the saddle point. 

This statement follows directly from the construction of the energy-dependent temperature (\ref{Eq_T_vs_E}), where $T(E) \to T_{\rm room}$ for $E \simeq \Delta E$. Due to this behaviour of $T(E)$, for one-step MCh probabilities (\ref{Eq_p_MCprob}) near the saddle point we have $p_{i \to i+1}^{\rm EDT} = p_{i \to i+1}^{T = Const}$ (and the same applies for $q$'s). Hence, as long as $b_{\rm cool}$ is large enough to allow $w_{i \to i+1}$ (computed from the matrix ${\bf \hat E}^{(i+1)}$) to reach its steady-state value for $T = T_{\rm room}$ in the saddle point region, in this region we should obtain $w_{i \to i+1}^{\rm EDT} = w_{i \to i+1}^{\rm CT}$. In the FFS method, probabilities ${w}$ are obtained from LD simulations, which 'feel' at each time integration step only the local temperature, so these probabilities should also be equal in the saddle-point region for $T = Const$ and $T(E)$ cases.

Corresponding ratios $w_{i \to i+1}^{\rm EDT}/w_{i \to i+1}^{\rm CT}$ are plotted in Fig. \ref{Fig_Prob_w_ratios_EDT_to_Tc} for the same system as in Fig. \ref{Fig_prob_FFS_vs_MCh_Tconst} and \ref{Fig_prob_FFS_vs_MCh_EDT}. It can be clearly seen that these ratios for FFS and MCh methods agree very well for all interface energies - as well far below the saddle point (where these ratios are governed only by the local temperatures) as in the saddle point region, where the dynamics of a real system plays a decisive role in the FFS method. This means that the ratio of the probability products (\ref{Eq_ratio_probprod}) required for the evaluation of the switching time in our energy-dependent temperature concept can be computed using the Matrix chain method. As stated above, this computation is very fast, involving only a few matrix multiplications to obtain the limit $\hat{\bf E}^{(i+1)} = \lim_{k \to \infty} (\hat{\bf P}^{(i+1)})^k$. Moreover, the ratio $r$ (\ref{Eq_ratio_probprod}) computed this way depends only on the function $T(E)$ and thus can be evaluated  for any system with the given barrier $\Delta E$ once and for all. 

Hence our algorithm for the switching time evaluation requires only numerical simulations of transitions over the barrier for the studied system with the {\it energy-dependent} temperature. This means that we have to collect only a sufficiently accurate statistics of transitions over the effective barrier with the height 
$\Delta E_{\rm eff} = \simeq \Delta E - E_{cool} = b_{\rm room} k_B T$. Corresponding simulation time is not only accessible for the direct LD modelling, but should be approximately {\it independent} on the height of the actual barrier $\Delta E$ - in strong contrast both to standard LD simulations (where simulation time 
$t_{\rm sim} \sim \exp(\Delta E / k_B T$) and FFS methods, where $t_{\rm sim} \sim \Delta E$.

Moreover, our method does suffer from the instability problem arsing in FFS and EnB algorithms due to the presence of the product of numerically computed transition probabilities, as explained in details in the Introduction. Computation of the probability product ratio $r$ (\ref{Eq_ratio_probprod}) in our method is error-free, so that this instability is completely absent.

Summarizing, our algorithm consists of the following stages:

(1) Divide the path between the basins ${\bf A}$ and ${\bf B}$ into $N$ states with the energy differences $\delta E$ between them.

(2) Set the energy-dependent temperature (EDT) (\ref{Eq_T_vs_E}).

(3) Using this $T(E)$ dependence, assign the one-step hopping probabilities $\{p_i\}$ and $\{q_i\}$ between the states according to (\ref{Eq_p_MCprob}) and (\ref{Eq_q_MCprob}).

(4) For each state $i$, build the transition matrix $\hat{\bf P}^{i+1}$ given by (\ref{Eq_One-step_trans_mtr_partial}) for the corresponding Markov chain.

(5) Compute the total EDT transition probabilities as matrix elements 
$w_{i \to i+1}^{\rm EDT} = E_{i,i+1}^{(i+1)}$, where 
$\hat{\bf E}^{(i+1)} = \lim_{k \to \infty} (\hat{\bf P}^{(i+1)})^k$ 

(6) Repeat the steps (3)-(5) for the constant temperature $T = T_{\rm room}$ to obtain the probabilities $w_{i \to i+1}^{\rm CT}$.

(7) Perform LD simulations for the EDT case and compute the EDT switching time 
$\tau_{\rm sw}^{\rm EDT}$ in a standard way.

(8) Perform LD simulations for $T = T_{\rm room}$ and $T = T_{\rm lrg}$ to compute the corresponding fluxes $\Phi_0^{\rm CT}$ and $\Phi_0^{\rm EDT}$ out of the basin {\bf A}.

(9) Compute the real switching time (for the constant temperature) $\tau_{\rm sw}^{\rm CT}$ according to (\ref{Eq_tau_sw_real}).

We emphasize once more that {\it the only really time-consuming step} in this algorithm is the procedure (7), where an accurate statistics of the switching events should be collected.

\section{Physical results and comparison of EDT with FFS}
\label{Sec_EDT_vs_FFS}

To demonstrate the high accuracy of our algorithm and to quantitatively compare the simulation time for determination of the switching rate in our EDT paradigm with the corresponding time required by FFS, we have simulated with both methods the same series of macrospins with the biaxial anisotropy as analysed in \cite{Semenova_Berkov_2020}, i.e. the macrospins with magnetic parameters as for Permalloy (magnetization $M = 800\, {\rm G}$, damping $\lambda = 0.01$) and demagnetizing factors of flat nanoellipses with the thickness $h = 3\, {\rm nm}$, short axis $a = 40 \, {\rm nm}$ and long axes $b$ varying from 50 to 100 nm; corresponding energy barriers are in the range $9 \le \Delta E / k_B T \le 60$.

\begin{figure}[htb]
\includegraphics[width=80mm]{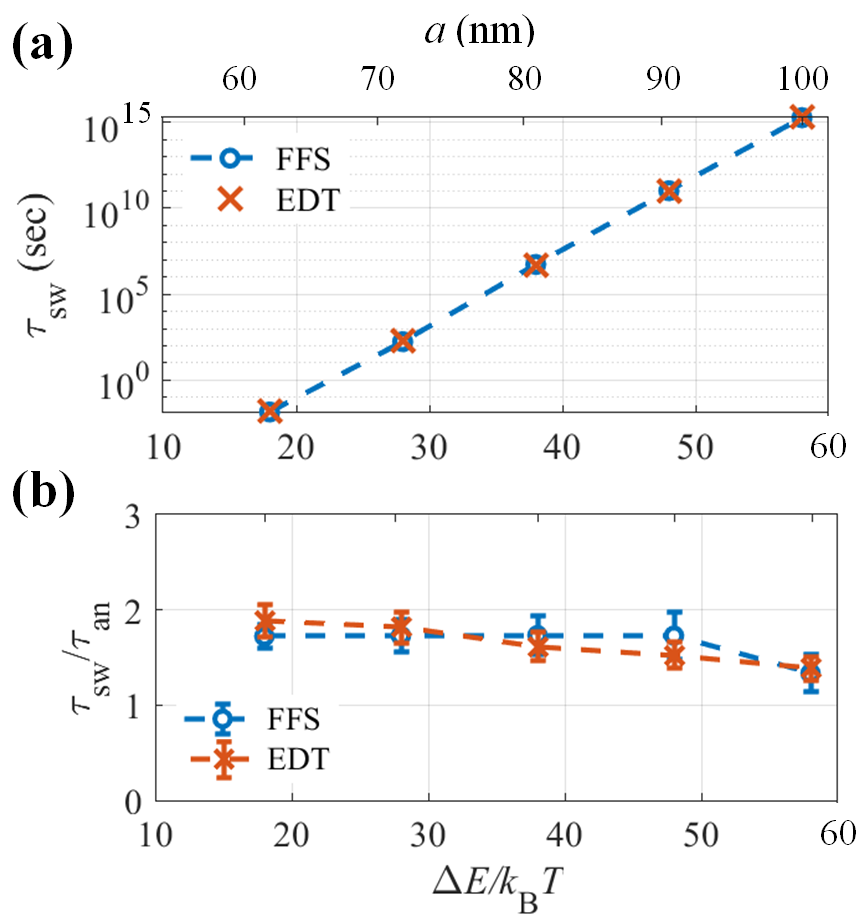}
\caption{Switching times computed by the standard ($T = Const$) FFS method (red open circles)  compared to the same times obtained via EDT. An excellent agreement between both methods is clearly demonstrated}.
\label{Fig_Tsw_EDT_FFS}
\end{figure}

Switching times for these macrospins cover approximately 20 orders of magnitude as shown in Fig. \ref{Fig_Tsw_EDT_FFS}(a). Our method demonstrates an excellent agreement with FFS simulations in the whole range of energy barriers. This agreement can be seen especially well in Fig. \ref{Fig_Tsw_EDT_FFS}(b), where FFS and EDT switching times obtained numerically are plotted as ratios between them and the analytical result obtained in \cite{Semenova_Berkov_2020} for the same macrospins.

\begin{figure}[htb]
\includegraphics[width=80mm]{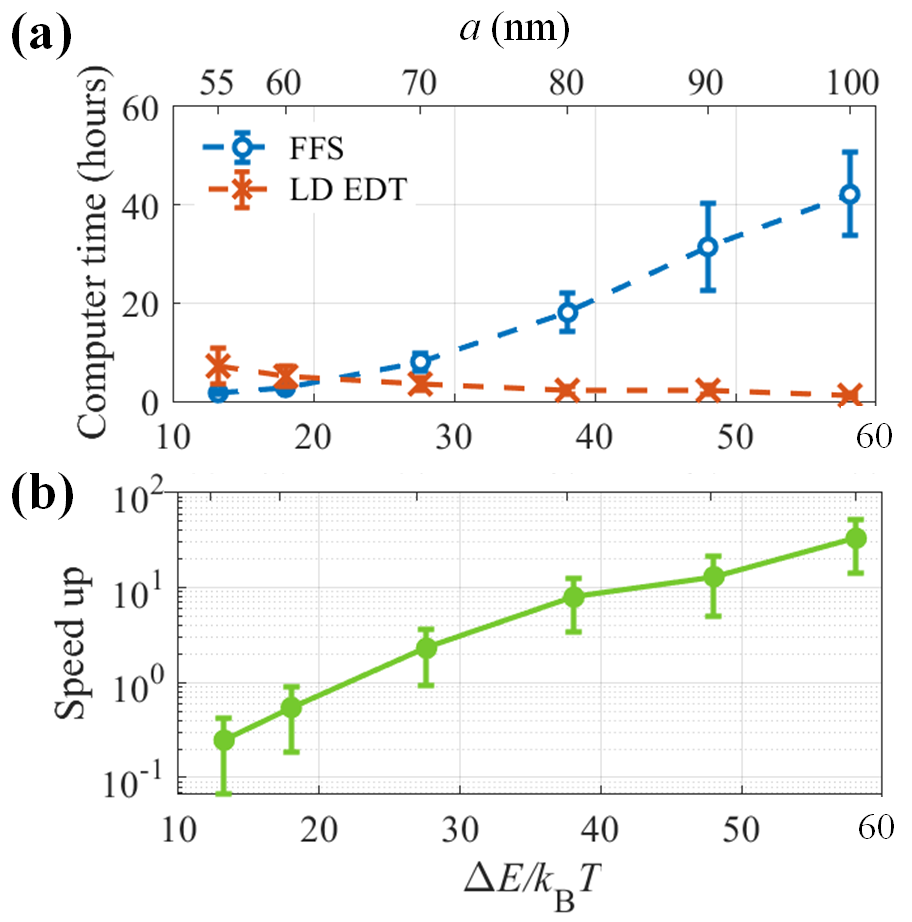}
\caption{Speed up. $\Delta t = 0.001 $, $5\% $ of accuracy (a) Calculation time for FFS and LD EDT ($b_{\rm cool} = 7$) methods; (b) speed up LD EDT vs. FFS}
\label{Fig_SpeedUp_FFS_vs_EDT}
\end{figure}

Finally, to compare the performance of our algorithm and the FFS method, we have determined simulation times required to compute the switching time with the relative accuracy $\epsilon = 5\%$ with both methods. Corresponding result plotted in Fig. \ref{Fig_SpeedUp_FFS_vs_EDT} confirm our conclusions drawn above. Namely, the FFS simulation time growths approximately linearly with the barrier height $\Delta E$, because the time required to compute each probability $w_i$ is approximately the same for each interface, and the required number of interfaces growth linearly with $\Delta E$. For our EDT algorithm, simulation time even decreases somewhat when the barrier increases, because for higher barriers the temperature $T_{\rm lrg}$ should be higher to ensure the same values of the probability $p(E_{\rm cool})$ (see Sec. \ref{Subsec_Main_idea}), so that the number of transitions over the barrier per unit time in EDT-LD simulations also increases. Thus we need smaller simulation time to obtain the statistics of the same quality.

The speedup of the EDT algorithm compared to FFS is shown in Fig. \ref{Fig_SpeedUp_FFS_vs_EDT}(b): the break point is achieved already for a very moderate barrier $\Delta E / k_B T  \approx 20$, and for the highest studied value $\Delta E / k_B T \approx 60$ our method is more the 40x faster than FFS.

%
\section{Conclusion}
\label{Sec_Conclusion}

In this paper we have introduced the concept of the energy-dependent temperature (EDT), which allows to simulate transitions over {\it arbitrary high} energy barriers by {\it single-stage} Langevin dynamics simulations. Our method has been verified on the example of a biaxial magnetic macrospin - the system with two energy minima and two equivalent saddle points - where our results agree very well with switching times obtained via the forward flux sampling (FFS). We have shown that the computation time for the EDT-based LD simulations {\it does not increase} with the energy barrier height, in contrast to FFS and other 'climbing' methods, thus providing a unique possibility to simulate transitions over any barrier with a very moderate numerical effort. The speedup of our LD-EDT method in comparison with the (strongly optimized) FFS simulation achieves 40x for the energy barrier of $\approx 60 k_B T$. Further, the presented EDT-LD algorithm does not require the evaluation of the product of a large number of conditional probabilities for transitions between subsequent interfaces as in FFS and related methods (like 'energy bounce' etc \cite{Semenova_Berkov_2020}) and thus does not suffer from the stability problem arising due to this procedure in presence of any systematic error occurs by the computation of these probabilities. 

\section{Acknowledgment}
\label{Sec_Acknow}
Financial support of the Deutsche Forschungsgemeinschaft (German Research Society), DFG-project BE 2464/18-1 is greatly acknowledged.

\bibliographystyle{apsrev4-1}
\bibliography{references}

\begin{thebibliography}{28}%
\makeatletter
\providecommand \@ifxundefined [1]{%
 \@ifx{#1\undefined}
}%
\providecommand \@ifnum [1]{%
 \ifnum #1\expandafter \@firstoftwo
 \else \expandafter \@secondoftwo
 \fi
}%
\providecommand \@ifx [1]{%
 \ifx #1\expandafter \@firstoftwo
 \else \expandafter \@secondoftwo
 \fi
}%
\providecommand \natexlab [1]{#1}%
\providecommand \enquote  [1]{``#1''}%
\providecommand \bibnamefont  [1]{#1}%
\providecommand \bibfnamefont [1]{#1}%
\providecommand \citenamefont [1]{#1}%
\providecommand \href@noop [0]{\@secondoftwo}%
\providecommand \href [0]{\begingroup \@sanitize@url \@href}%
\providecommand \@href[1]{\@@startlink{#1}\@@href}%
\providecommand \@@href[1]{\endgroup#1\@@endlink}%
\providecommand \@sanitize@url [0]{\catcode `\\12\catcode `\$12\catcode
  `\&12\catcode `\#12\catcode `\^12\catcode `\_12\catcode `\%12\relax}%
\providecommand \@@startlink[1]{}%
\providecommand \@@endlink[0]{}%
\providecommand \url  [0]{\begingroup\@sanitize@url \@url }%
\providecommand \@url [1]{\endgroup\@href {#1}{\urlprefix }}%
\providecommand \urlprefix  [0]{URL }%
\providecommand \Eprint [0]{\href }%
\providecommand \doibase [0]{http://dx.doi.org/}%
\providecommand \selectlanguage [0]{\@gobble}%
\providecommand \bibinfo  [0]{\@secondoftwo}%
\providecommand \bibfield  [0]{\@secondoftwo}%
\providecommand \translation [1]{[#1]}%
\providecommand \BibitemOpen [0]{}%
\providecommand \bibitemStop [0]{}%
\providecommand \bibitemNoStop [0]{.\EOS\space}%
\providecommand \EOS [0]{\spacefactor3000\relax}%
\providecommand \BibitemShut  [1]{\csname bibitem#1\endcsname}%
\let\auto@bib@innerbib\@empty
\bibitem [{\citenamefont {Haenggi}\ \emph {et~al.}(1990)\citenamefont
  {Haenggi}, \citenamefont {Talkner},\ and\ \citenamefont
  {Borkovec}}]{Haenggi_RevModPhys_1990}%
  \BibitemOpen
  \bibfield  {author} {\bibinfo {author} {\bibfnamefont {P.}~\bibnamefont
  {Haenggi}}, \bibinfo {author} {\bibfnamefont {P.}~\bibnamefont {Talkner}}, \
  and\ \bibinfo {author} {\bibfnamefont {M.}~\bibnamefont {Borkovec}},\ }\href
  {\doibase 10.1103/RevModPhys.62.251} {\bibfield  {journal} {\bibinfo
  {journal} {Rev. Mod. Phys.}\ }\textbf {\bibinfo {volume} {62}},\ \bibinfo
  {pages} {251} (\bibinfo {year} {1990})}\BibitemShut {NoStop}%
\bibitem [{\citenamefont {Jonsson}\ \emph {et~al.}(1998)\citenamefont
  {Jonsson}, \citenamefont {Mills},\ and\ \citenamefont
  {Jacobsen}}]{Jonsson_1998}%
  \BibitemOpen
  \bibfield  {author} {\bibinfo {author} {\bibfnamefont {H.}~\bibnamefont
  {Jonsson}}, \bibinfo {author} {\bibfnamefont {G.}~\bibnamefont {Mills}}, \
  and\ \bibinfo {author} {\bibfnamefont {K.}~\bibnamefont {Jacobsen}},\
  }\enquote {\bibinfo {title} {Nudged elastic band method for ﬁnding minimum
  energy paths of transitions},}\ in\ \href {\doibase
  10.1142/9789812839664_0016} {\emph {\bibinfo {booktitle} {Classical and
  Quantum Dynamics in Condensed Phase Simulations}}}\ (\bibinfo  {publisher}
  {World Scientiﬁc, Singapore},\ \bibinfo {year} {1998})\ Chap.~\bibinfo
  {chapter} {16}, pp.\ \bibinfo {pages} {385--404}\BibitemShut {NoStop}%
\bibitem [{\citenamefont {E}\ \emph {et~al.}(2002)\citenamefont {E},
  \citenamefont {Ren},\ and\ \citenamefont {Vanden-Eijnden}}]{E_2002}%
  \BibitemOpen
  \bibfield  {author} {\bibinfo {author} {\bibfnamefont {W.}~\bibnamefont {E}},
  \bibinfo {author} {\bibfnamefont {W.}~\bibnamefont {Ren}}, \ and\ \bibinfo
  {author} {\bibfnamefont {E.}~\bibnamefont {Vanden-Eijnden}},\ }\href@noop {}
  {\bibfield  {journal} {\bibinfo  {journal} {Phys. Rev. B}\ }\textbf {\bibinfo
  {volume} {60}},\ \bibinfo {pages} {052301} (\bibinfo {year}
  {2002})}\BibitemShut {NoStop}%
\bibitem [{\citenamefont {Berkov}(2007)}]{Berkov_HMMM_2007}%
  \BibitemOpen
  \bibfield  {author} {\bibinfo {author} {\bibfnamefont {D.}~\bibnamefont
  {Berkov}},\ }\enquote {\bibinfo {title} {Magnetization dynamics including
  thermal fluctuations},}\ in\ \href@noop {} {\emph {\bibinfo {booktitle}
  {Handbook of Magnetism and Advanced Magnetic Materials}}},\ Vol.~\bibinfo
  {volume} {2},\ \bibinfo {editor} {edited by\ \bibinfo {editor} {\bibfnamefont
  {H.}~\bibnamefont {Kronmu\"ller}}\ and\ \bibinfo {editor} {\bibfnamefont
  {S.}~\bibnamefont {Parkin}}}\ (\bibinfo  {publisher} {John Wiley \& Sons
  Ltd},\ \bibinfo {year} {2007})\ Chap.~\bibinfo {chapter} {4}, pp.\ \bibinfo
  {pages} {795--823}\BibitemShut {NoStop}%
\bibitem [{\citenamefont {Onsager}\ and\ \citenamefont
  {Machlup}(1953)}]{Onsager_1953}%
  \BibitemOpen
  \bibfield  {author} {\bibinfo {author} {\bibfnamefont {L.}~\bibnamefont
  {Onsager}}\ and\ \bibinfo {author} {\bibfnamefont {S.}~\bibnamefont
  {Machlup}},\ }\href@noop {} {\bibfield  {journal} {\bibinfo  {journal} {Phys.
  Rev.}\ }\textbf {\bibinfo {volume} {91}},\ \bibinfo {pages} {1505} (\bibinfo
  {year} {1953})}\BibitemShut {NoStop}%
\bibitem [{\citenamefont {Berkov}(1998)}]{Berkov_1998}%
  \BibitemOpen
  \bibfield  {author} {\bibinfo {author} {\bibfnamefont {D.}~\bibnamefont
  {Berkov}},\ }\href@noop {} {\bibfield  {journal} {\bibinfo  {journal} {J.
  Magn. Magn. Mat.}\ }\textbf {\bibinfo {volume} {186}},\ \bibinfo {pages}
  {199} (\bibinfo {year} {1998})}\BibitemShut {NoStop}%
\bibitem [{\citenamefont {Braun}(1994)}]{Braun_1994}%
  \BibitemOpen
  \bibfield  {author} {\bibinfo {author} {\bibfnamefont {H.-B.}\ \bibnamefont
  {Braun}},\ }\href@noop {} {\bibfield  {journal} {\bibinfo  {journal} {J.
  Appl. Phys.}\ }\textbf {\bibinfo {volume} {76}},\ \bibinfo {pages} {6310}
  (\bibinfo {year} {1994})}\BibitemShut {NoStop}%
\bibitem [{\citenamefont {Fiedler}\ \emph {et~al.}(2012)\citenamefont
  {Fiedler}, \citenamefont {Fidler}, \citenamefont {Lee}, \citenamefont
  {Schrefl}, \citenamefont {Stamps}, \citenamefont {Braun},\ and\ \citenamefont
  {Suess}}]{Fiedler_2012}%
  \BibitemOpen
  \bibfield  {author} {\bibinfo {author} {\bibfnamefont {G.}~\bibnamefont
  {Fiedler}}, \bibinfo {author} {\bibfnamefont {J.}~\bibnamefont {Fidler}},
  \bibinfo {author} {\bibfnamefont {J.}~\bibnamefont {Lee}}, \bibinfo {author}
  {\bibfnamefont {T.}~\bibnamefont {Schrefl}}, \bibinfo {author} {\bibfnamefont
  {R.~L.}\ \bibnamefont {Stamps}}, \bibinfo {author} {\bibfnamefont
  {H.}~\bibnamefont {Braun}}, \ and\ \bibinfo {author} {\bibfnamefont
  {D.}~\bibnamefont {Suess}},\ }\href {\doibase 10.1063/1.4712033} {\bibfield
  {journal} {\bibinfo  {journal} {J. Appl. Phys.}\ }\textbf {\bibinfo {volume}
  {111}},\ \bibinfo {pages} {093917} (\bibinfo {year} {2012})}\BibitemShut
  {NoStop}%
\bibitem [{\citenamefont {Kramers}(1940)}]{Kramers_1940}%
  \BibitemOpen
  \bibfield  {author} {\bibinfo {author} {\bibfnamefont {H.}~\bibnamefont
  {Kramers}},\ }\href@noop {} {\bibfield  {journal} {\bibinfo  {journal}
  {Physica}\ }\textbf {\bibinfo {volume} {7}},\ \bibinfo {pages} {284}
  (\bibinfo {year} {1940})}\BibitemShut {NoStop}%
\bibitem [{\citenamefont {Coffey}\ and\ \citenamefont
  {Kalmykov}(2012)}]{Coffey_Kalmykov_2012}%
  \BibitemOpen
  \bibfield  {author} {\bibinfo {author} {\bibfnamefont {W.}~\bibnamefont
  {Coffey}}\ and\ \bibinfo {author} {\bibfnamefont {Y.}~\bibnamefont
  {Kalmykov}},\ }\href@noop {} {\bibfield  {journal} {\bibinfo  {journal} {J.
  Appl. Phys.}\ }\textbf {\bibinfo {volume} {112}},\ \bibinfo {pages} {121301}
  (\bibinfo {year} {2012})}\BibitemShut {NoStop}%
\bibitem [{\citenamefont {Mel'nikov}\ and\ \citenamefont
  {Meshkov}(1986)}]{Melnikov_1986}%
  \BibitemOpen
  \bibfield  {author} {\bibinfo {author} {\bibfnamefont {V.}~\bibnamefont
  {Mel'nikov}}\ and\ \bibinfo {author} {\bibfnamefont {S.}~\bibnamefont
  {Meshkov}},\ }\href@noop {} {\bibfield  {journal} {\bibinfo  {journal} {J.
  Chem. Phys.}\ }\textbf {\bibinfo {volume} {85}},\ \bibinfo {pages} {1018}
  (\bibinfo {year} {1986})}\BibitemShut {NoStop}%
\bibitem [{\citenamefont {Brown~Jr}(1979)}]{Brown_1979}%
  \BibitemOpen
  \bibfield  {author} {\bibinfo {author} {\bibfnamefont {W.~F.}\ \bibnamefont
  {Brown~Jr}},\ }\href@noop {} {\bibfield  {journal} {\bibinfo  {journal} {IEEE
  Trans. Magn.}\ }\textbf {\bibinfo {volume} {MAG-15}},\ \bibinfo {pages}
  {1196} (\bibinfo {year} {1979})}\BibitemShut {NoStop}%
\bibitem [{\citenamefont {Klik}\ and\ \citenamefont
  {Gunther}(1990)}]{Klik_1990}%
  \BibitemOpen
  \bibfield  {author} {\bibinfo {author} {\bibfnamefont {I.}~\bibnamefont
  {Klik}}\ and\ \bibinfo {author} {\bibfnamefont {L.}~\bibnamefont {Gunther}},\
  }\href@noop {} {\bibfield  {journal} {\bibinfo  {journal} {J. Stat. Phys.}\
  }\textbf {\bibinfo {volume} {60}},\ \bibinfo {pages} {473} (\bibinfo {year}
  {1990})}\BibitemShut {NoStop}%
\bibitem [{\citenamefont {Coffey}\ \emph {et~al.}(2001)\citenamefont {Coffey},
  \citenamefont {Garanin},\ and\ \citenamefont {McCarthy}}]{Coffey_2001}%
  \BibitemOpen
  \bibfield  {author} {\bibinfo {author} {\bibfnamefont {W.~T.}\ \bibnamefont
  {Coffey}}, \bibinfo {author} {\bibfnamefont {D.~A.}\ \bibnamefont {Garanin}},
  \ and\ \bibinfo {author} {\bibfnamefont {D.~J.}\ \bibnamefont {McCarthy}},\
  }\href@noop {} {\bibfield  {journal} {\bibinfo  {journal} {Adv. Chem. Phys.}\
  }\textbf {\bibinfo {volume} {117}},\ \bibinfo {pages} {483} (\bibinfo {year}
  {2001})}\BibitemShut {NoStop}%
\bibitem [{\citenamefont {D\'ejardin}\ \emph {et~al.}(2001)\citenamefont
  {D\'ejardin}, \citenamefont {Crothers}, \citenamefont {Coffey},\ and\
  \citenamefont {McCarthy}}]{Dejardin_2001}%
  \BibitemOpen
  \bibfield  {author} {\bibinfo {author} {\bibfnamefont {P.~M.}\ \bibnamefont
  {D\'ejardin}}, \bibinfo {author} {\bibfnamefont {D.~S.~F.}\ \bibnamefont
  {Crothers}}, \bibinfo {author} {\bibfnamefont {W.~T.}\ \bibnamefont
  {Coffey}}, \ and\ \bibinfo {author} {\bibfnamefont {D.~J.}\ \bibnamefont
  {McCarthy}},\ }\href@noop {} {\bibfield  {journal} {\bibinfo  {journal}
  {Phys. Rev. E}\ }\textbf {\bibinfo {volume} {63}},\ \bibinfo {pages} {021102}
  (\bibinfo {year} {2001})}\BibitemShut {NoStop}%
\bibitem [{\citenamefont {Hubert}(1998)}]{Hubert_book_1998}%
  \BibitemOpen
  \bibfield  {author} {\bibinfo {author} {\bibfnamefont {A.}~\bibnamefont
  {Hubert}},\ }\href@noop {} {\emph {\bibinfo {title} {Magnetic Domains: The
  Analysis of Magnetic Microstructures}}}\ (\bibinfo  {publisher}
  {Springer-Verlag, Berlin},\ \bibinfo {year} {1998})\BibitemShut {NoStop}%
\bibitem [{\citenamefont {Allen}\ \emph {et~al.}(2005)\citenamefont {Allen},
  \citenamefont {Warren},\ and\ \citenamefont {ten Wolde}}]{Allen_2005}%
  \BibitemOpen
  \bibfield  {author} {\bibinfo {author} {\bibfnamefont {R.~J.}\ \bibnamefont
  {Allen}}, \bibinfo {author} {\bibfnamefont {P.~B.}\ \bibnamefont {Warren}}, \
  and\ \bibinfo {author} {\bibfnamefont {P.~R.}\ \bibnamefont {ten Wolde}},\
  }\href {\doibase 10.1103/PhysRevLett.94.018104} {\bibfield  {journal}
  {\bibinfo  {journal} {Phys. Rev. Lett.}\ }\textbf {\bibinfo {volume} {94}},\
  \bibinfo {pages} {018104} (\bibinfo {year} {2005})}\BibitemShut {NoStop}%
\bibitem [{\citenamefont {Allen}\ \emph {et~al.}(2006)\citenamefont {Allen},
  \citenamefont {Frenkel},\ and\ \citenamefont {ten Wolde}}]{Allen_2006b}%
  \BibitemOpen
  \bibfield  {author} {\bibinfo {author} {\bibfnamefont {R.~J.}\ \bibnamefont
  {Allen}}, \bibinfo {author} {\bibfnamefont {D.}~\bibnamefont {Frenkel}}, \
  and\ \bibinfo {author} {\bibfnamefont {P.~R.}\ \bibnamefont {ten Wolde}},\
  }\href {\doibase 10.1063/1.2198827} {\bibfield  {journal} {\bibinfo
  {journal} {J. Chem. Phys.}\ }\textbf {\bibinfo {volume} {124}},\ \bibinfo
  {pages} {194111} (\bibinfo {year} {2006})}\BibitemShut {NoStop}%
\bibitem [{\citenamefont {Borrero}\ and\ \citenamefont
  {Escobedo}(2008)}]{Borrero_2008}%
  \BibitemOpen
  \bibfield  {author} {\bibinfo {author} {\bibfnamefont {E.~E.}\ \bibnamefont
  {Borrero}}\ and\ \bibinfo {author} {\bibfnamefont {F.~A.}\ \bibnamefont
  {Escobedo}},\ }\href {\doibase 10.1063/1.2953325} {\bibfield  {journal}
  {\bibinfo  {journal} {J. Chem. Phys.}\ }\textbf {\bibinfo {volume} {129}},\
  \bibinfo {pages} {024115} (\bibinfo {year} {2008})}\BibitemShut {NoStop}%
\bibitem [{\citenamefont {Allen}\ \emph {et~al.}(2009)\citenamefont {Allen},
  \citenamefont {Valeriani},\ and\ \citenamefont {ten Wolde}}]{Allen_2009}%
  \BibitemOpen
  \bibfield  {author} {\bibinfo {author} {\bibfnamefont {R.}~\bibnamefont
  {Allen}}, \bibinfo {author} {\bibfnamefont {C.}~\bibnamefont {Valeriani}}, \
  and\ \bibinfo {author} {\bibfnamefont {P.~R.}\ \bibnamefont {ten Wolde}},\
  }\href {\doibase 10.1088/0953-8984/21/46/463102} {\bibfield  {journal}
  {\bibinfo  {journal} {J. Phys.: Cond. Matt.}\ }\textbf {\bibinfo {volume}
  {21}},\ \bibinfo {pages} {463102} (\bibinfo {year} {2009})}\BibitemShut
  {NoStop}%
\bibitem [{\citenamefont {Vogler}\ \emph {et~al.}(2013)\citenamefont {Vogler},
  \citenamefont {Bruckner}, \citenamefont {Bergmair}, \citenamefont {Huber},
  \citenamefont {Suess},\ and\ \citenamefont {Dellago}}]{Vogler_2013}%
  \BibitemOpen
  \bibfield  {author} {\bibinfo {author} {\bibfnamefont {C.}~\bibnamefont
  {Vogler}}, \bibinfo {author} {\bibfnamefont {F.}~\bibnamefont {Bruckner}},
  \bibinfo {author} {\bibfnamefont {B.}~\bibnamefont {Bergmair}}, \bibinfo
  {author} {\bibfnamefont {T.}~\bibnamefont {Huber}}, \bibinfo {author}
  {\bibfnamefont {D.}~\bibnamefont {Suess}}, \ and\ \bibinfo {author}
  {\bibfnamefont {C.}~\bibnamefont {Dellago}},\ }\href@noop {} {\bibfield
  {journal} {\bibinfo  {journal} {Phys. Rev. B}\ }\textbf {\bibinfo {volume}
  {88}},\ \bibinfo {pages} {134409} (\bibinfo {year} {2013})}\BibitemShut
  {NoStop}%
\bibitem [{\citenamefont {Vogler}\ \emph {et~al.}(2015)\citenamefont {Vogler},
  \citenamefont {Bruckner}, \citenamefont {Suess},\ and\ \citenamefont
  {Dellago}}]{Vogler_2015}%
  \BibitemOpen
  \bibfield  {author} {\bibinfo {author} {\bibfnamefont {C.}~\bibnamefont
  {Vogler}}, \bibinfo {author} {\bibfnamefont {F.}~\bibnamefont {Bruckner}},
  \bibinfo {author} {\bibfnamefont {D.}~\bibnamefont {Suess}}, \ and\ \bibinfo
  {author} {\bibfnamefont {C.}~\bibnamefont {Dellago}},\ }\href {\doibase
  10.1063/1.4918902} {\bibfield  {journal} {\bibinfo  {journal} {J. Appl.
  Phys.}\ }\textbf {\bibinfo {volume} {117}},\ \bibinfo {pages} {163907}
  (\bibinfo {year} {2015})}\BibitemShut {NoStop}%
\bibitem [{\citenamefont {Desplat}\ and\ \citenamefont
  {Kim}(2020)}]{Desplat_2020b}%
  \BibitemOpen
  \bibfield  {author} {\bibinfo {author} {\bibfnamefont {L.}~\bibnamefont
  {Desplat}}\ and\ \bibinfo {author} {\bibfnamefont {J.-V.}\ \bibnamefont
  {Kim}},\ }\href {\doibase 10.1103/PhysRevApplied.14.064064} {\bibfield
  {journal} {\bibinfo  {journal} {Physical Review Applied}\ }\textbf {\bibinfo
  {volume} {14}},\ \bibinfo {pages} {064064} (\bibinfo {year}
  {2020})}\BibitemShut {NoStop}%
\bibitem [{\citenamefont {Desplat}\ \emph {et~al.}(2020)\citenamefont
  {Desplat}, \citenamefont {Vogler}, \citenamefont {Kim}, \citenamefont
  {Stamps},\ and\ \citenamefont {Suess}}]{Desplat_2020a}%
  \BibitemOpen
  \bibfield  {author} {\bibinfo {author} {\bibfnamefont {L.}~\bibnamefont
  {Desplat}}, \bibinfo {author} {\bibfnamefont {C.}~\bibnamefont {Vogler}},
  \bibinfo {author} {\bibfnamefont {J.-V.}\ \bibnamefont {Kim}}, \bibinfo
  {author} {\bibfnamefont {R.~L.}\ \bibnamefont {Stamps}}, \ and\ \bibinfo
  {author} {\bibfnamefont {D.}~\bibnamefont {Suess}},\ }\href@noop {}
  {\bibfield  {journal} {\bibinfo  {journal} {Phys. Rev. B}\ }\textbf {\bibinfo
  {volume} {101}},\ \bibinfo {pages} {060403(R)} (\bibinfo {year}
  {2020})}\BibitemShut {NoStop}%
\bibitem [{\citenamefont {Semenova}\ \emph {et~al.}(2020)\citenamefont
  {Semenova}, \citenamefont {Berkov},\ and\ \citenamefont
  {Gorn}}]{Semenova_Berkov_2020}%
  \BibitemOpen
  \bibfield  {author} {\bibinfo {author} {\bibfnamefont {E.~K.}\ \bibnamefont
  {Semenova}}, \bibinfo {author} {\bibfnamefont {D.~V.}\ \bibnamefont
  {Berkov}}, \ and\ \bibinfo {author} {\bibfnamefont {N.~L.}\ \bibnamefont
  {Gorn}},\ }\href {\doibase 10.1103/PhysRevB.102.144419} {\bibfield  {journal}
  {\bibinfo  {journal} {Phys. Rev. B}\ }\textbf {\bibinfo {volume} {102}},\
  \bibinfo {pages} {144419} (\bibinfo {year} {2020})}\BibitemShut {NoStop}%
\bibitem [{\citenamefont {Wang}\ and\ \citenamefont
  {Visscher}(2006)}]{Wang_Visscher_2006}%
  \BibitemOpen
  \bibfield  {author} {\bibinfo {author} {\bibfnamefont {S.}~\bibnamefont
  {Wang}}\ and\ \bibinfo {author} {\bibfnamefont {P.}~\bibnamefont
  {Visscher}},\ }\href {\doibase 10.1063/1.2176868} {\bibfield  {journal}
  {\bibinfo  {journal} {J. Appl. Phys.}\ }\textbf {\bibinfo {volume} {99}},\
  \bibinfo {pages} {08G106} (\bibinfo {year} {2006})}\BibitemShut {NoStop}%
\bibitem [{\citenamefont {Howard}(2012)}]{Howard_DynProbSys_book}%
  \BibitemOpen
  \bibfield  {author} {\bibinfo {author} {\bibfnamefont {R.}~\bibnamefont
  {Howard}},\ }\href@noop {} {\emph {\bibinfo {title} {Dynamic Probabilistic
  Systems: Markov Models}}}\ (\bibinfo  {publisher} {J. Wiley \& Sons},\
  \bibinfo {year} {2012})\BibitemShut {NoStop}%
\bibitem [{\citenamefont {van Kampen}(1992)}]{vanKampen_book}%
  \BibitemOpen
  \bibfield  {author} {\bibinfo {author} {\bibfnamefont {N.}~\bibnamefont {van
  Kampen}},\ }\href@noop {} {\emph {\bibinfo {title} {Stochastic Processes in
  Physics and Chemistry}}}\ (\bibinfo  {publisher} {Elsevier Science},\
  \bibinfo {year} {1992})\BibitemShut {NoStop}%
\end{thebibliography}%

\end{document}